# Elastic deformation during dynamic force measurements in viscous fluids

Yumo Wang[†1], Georgia Pilkington[†1], Charles Dhong[1], and Joelle Frechette[*,1,2]

[1]Chemical and Biomolecular Engineering Department and [2]Hopkins Extreme Materials Institute, Johns Hopkins University, Baltimore MD 21218

* Corresponding Author, jfrechette@jhu.edu

[†] Equal contribution

**Abstract**

Understanding and harnessing the coupling between lubrication pressure and elasticity provides materials design strategies for applications such as adhesives, coatings, microsensors, and biomaterials. Elastic deformation of compliant solids caused by viscous forces can also occur during dynamic force measurements in instruments such as the surface forces apparatus (SFA) or the atomic force microscope (AFM). We briefly review hydrodynamic interactions in the presence of soft, deformable interfaces in the lubrication limit. More specifically, we consider the scenario of two surfaces approaching each other in a viscous fluid where one or both surfaces is deformable, which is also relevant to many force measurement systems. In this article the basic theoretical background of the elastohydrodynamic problem is detailed, followed by a discussion of experimental validation and considerations, especially for the role of elastic deformation on surface forces measurements. Finally, current challenges to our understanding of soft hydrodynamic interactions, such as the consideration of substrate layering, poroelasticity, viscoelasticity, surface heterogeneity, as well as their implications are discussed.

**Introduction**

Viscous forces caused by the relative movement of two surfaces in a viscous fluid can exert local pressures that can be sufficiently large to cause elastic deformation of the interacting materials (see Figure 1). This scenario is very common in the tribology of lubricated contacts, where deformation due to viscous forces in oils is a key mechanism to reduce friction and prevent wear (i.e. elastohydrodynamic lubrication or EHL, see Figure 1a).[1] Similarly, the presence of an elastic boundary during the drainage or infusion of fluid in a confined gap can lead to elastohydrodynamic deformation (EHD, see Figure 1b).

Both EHD and EHL play an important role in soft matter where materials such as gels, biological tissues, or elastomers (Young's modulus~10kPa-100MPa) can deform during motion in a fluid at relatively low velocities and viscosities.[2-5] For example, elastic deformation will affect the collision and rheological response of soft colloidal particles (Figure 1c (i) and (ii)), coalescence of bubbles/drops (Figure 1c (iii)), and has been hypothesized as a mechanism for shear thickening behavior.[6]. Studying the dynamic adhesion of cells or capsules attached to a wall is also accompanied by deformation, and the stored elastic energy greatly alters the detachment process (Figure 1d).[7] Similarly, insects and several vertebrates are also known to mediate contact between their soft toe pads and surfaces through liquid layers[8], therefore understanding their locomotion and detachment requires understanding how the toe pads and highly confined liquid layers interact. Finally, in joint cartilage, a lubricating layer of fluid and the squeeze out from a polymer network (weeping lubrication) is suspected to be responsible for the low apparent friction across joints.[9,10]

These deformations, in turn, can have a profound effect on the hydrodynamic interactions [11], such as those encountered during dynamic force measurements with the atomic force microscope (AFM)[12,13] or the surface forces apparatus (SFA) (see Figure 1b).[14,15] The SFA and AFM are well-suited to study the coupling between soft matter and hydrodynamic interactions, as demonstrated with droplets or bubbles[16-21](Figure 1c(iii)). More recently, these tools have been employed and adapted to study elastohydrodynamic deformation of compliant solids.[13-15] It is also important to consider the possibility of elastic deformation in the interpretation of surface forces measurements when one (or both) of the interacting surfaces has some degree of compliance (or reversely to know when to confidently ignore it). Neglecting to account for elastic deformation could lead to misinterpretation of the conservative surface forces (such as van der Waals or electrostatic) or slip behavior.

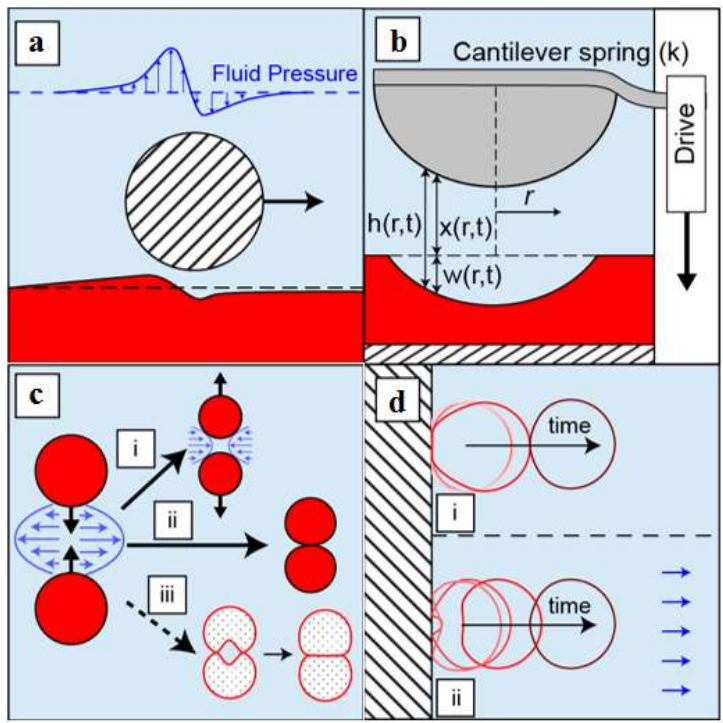

**Figure 1.** Hydrodynamic interactions in the presence of deformable materials. (a) Elastohydrodynamic lubrication: lift generated due to the sliding of a rigid cylinder along a soft material. (b) Elastohydrodynamic force measurement in the sphere-wall configuration with labeled variables. (Not to scale) (c) Rheological behavior of particles, bubbles, or drops colliding followed by either (i) rebound or (ii) sticking, or (iii) dimpling and then coalescence for bubbles/drops. (d) Adhesion and detachment of a soft or liquid-filled capsule, or cell, where the morphology and dynamics of detachment changes in the presence of quiescent (i) or external (axisymmetric) flow (ii). (d) is adapted from [22] with permission.

Due to its broad practical significance, a better understanding of the impact of deformation on hydrodynamic interactions and surface forces is necessary. Here we review studies, which in general have considered the classic geometry of a sphere approaching a plate at a constant drive velocity in a viscous fluid within the lubrication limit, and where one of the surfaces is supported by a force measuring spring (see Figure 1b). These are the experimental conditions encountered in typical surface forces measurements. This review is organized as follows: first we describe the theoretical framework and present experimental validations, focusing on direct force measurements. We then discuss current scientific challenges in describing or characterizing hydrodynamic interactions in the presence of elastic deformation. Finally, we provide broad guidelines to researchers working on surface forces measurements that highlight experimental conditions where elastic deformation might be present in dynamic force measurements with the AFM or SFA. Important and related topics not covered here include bouncing and rebounding of spheres[6](Fig.1c), jamming of deformable spheres[23], elastohydrodynamic lubrication (EHL), and rheological implications[24].

## 2. Physical Description

While there are many configurations where elastohydrodynamic deformation might be important (see Figure 1), the most common alignment encountered surface force measurements is the sphere-plane geometry. This configuration facilitates experimentation and analysis due to axisymmetry and by ensuring point contact. (Figure 1b). A geometrically equivalent configuration would be that of two nearby cross-cylinders which is used in the SFA.[25] The description of the elastohydrodynamic problem needs to address the coupling between fluid dynamics and surface deformations. The lubrication approximation with the no-slip boundary condition (Equation 1) describes drainage and infusion of liquid from a thin gap in the limit where the surface separation is much smaller than the radial length scale:

$$\frac{\partial h}{\partial t} = \frac{1}{12\mu r} \frac{\partial}{\partial r}\left(r h^3 \frac{\partial p}{\partial r}\right), \qquad [1]$$

where $h$ is surface separation, $t$ is time, $\mu$ is the viscosity for a Newtonian fluid and $p$ is fluid pressure. [26,27] In the absence of elastic deformation the fluid pressure distribution is readily obtained from Eq. 1 for a given initial condition, $h(r,0)$. The pressure can then be integrated radially to determine the total hydrodynamic force.

Surface deformation alters the fluid film thickness profile and shape of the interacting surfaces. This change in local fluid film thickness modifies the fluid pressure which, in turn, determines the deformation profile. The deformation profile, $w(r,t)$, can be obtained from linear elasticity for a spherical half-space, and is given by Equation 2:

$$w(r,t) = 4\theta \int_0^\infty f(y,t) \frac{y}{y+r} K\left[\frac{4ry}{(r+y)^2}\right] dy, \qquad [2]$$

where $\theta = \frac{1-v^2}{\pi E}$, $v$ is Poisson's ratio, $E$ is Young's modulus, $f$ is surface stress distribution, $y$ is a dummy variable used in integration, and $K$ is the complete elliptic integral of the first kind. In Eq.

2 the surface stress distribution comes from the fluid pressure, and due to the solid-fluid coupling, the lubrication equation needs to be solved simultaneously with the deformation profile. Usually a numerical solution is necessary, but analytical solutions can be obtained by making some approximations. For instance, Davis et al. [28] demonstrated that in the limit of small deformation (*w/x<0.05* in Figure 1b) a closed-form solution can be obtained by using the undeformed surface separation in the lubrication equation. This solution has served as a useful means of examining whether notable deformation is present in drainage flow.[25] A simple algorithm proposed for modeling half-space in the case of significant deformation is to treat the deformation as a non-adhesive contact (Hertz theory[29]), and to replace the contact pressure distribution with the fluid pressure obtained for a given gap thickness by the lubrication equation. [30]

A more general consideration than assuming a half-space is necessary to treat the case of hard surfaces with compliant coatings or more generally for layered materials. In this case, the deformation can be acquired by applying a sticky boundary condition on the rigid substrate, using fluid pressure as a surface condition, and then solve for linear elasticity theory.[31] This approach allows for the role of the coating thickness to be investigated directly. For instance, Leroy and Charlaix used this approach to characterize regimes where fluid viscosity or the elasticity of the soft materials dominates the force response when confined by an oscillating rigid indenter (see Figure 2a).[3] Using this method they demonstrated that the absolute modulus of soft coatings could be measured out of contact, and without measuring the contributions of the underlying substrate.

During surface forces measurements one surface is typically mounted on a spring (or cantilever). In this case, the surface separation is different from the displacement of the driving motor due to the deflection of the spring. The different contributions to the total force, $F_T$, from the hydrodynamic and conservative forces ($F_s$, e.g. van der Waals force, double layer force) can be considered independently[16] and this force balance for the sphere-wall geometry is given by Eq. 3:

$$F_T = kS(t) = \int_0^R 2\pi r p(r) dr + F_s, \qquad [3]$$

where *R* is the radius of spherical particle, and *S(t)* is the deflection of the cantilever with a spring constant *k* as a function of time.

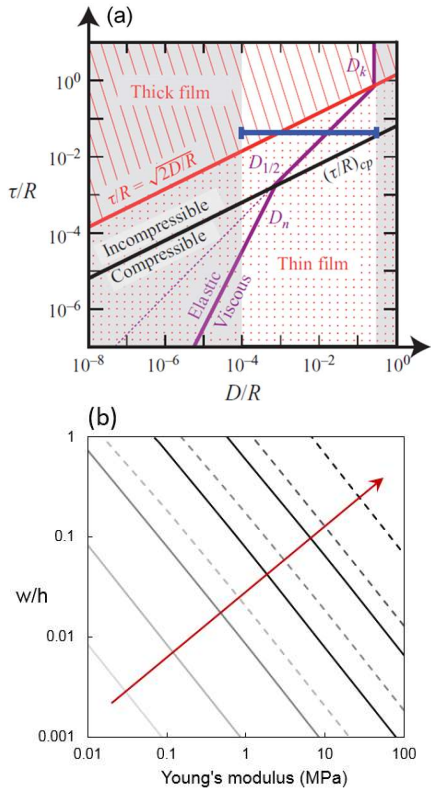

**Figure. 2** (a) Dynamic phase diagram for an elastic film of thickness $\tau$ with a Poisson's ratio $v$ =0.499. D is the surface separation, R represents the sphere radius. The continuous red line is the thick/thin film transition and the parallel black line is the compressible–incompressible transition, where the incompressible thin film domain lies between those lines. The purple lines are the elastic/viscous cross-over distances. The horizontal blue line is an example of experimental space which crosses several transitions. Figure is reprinted from [3] with permission (b) Calculated deformation (normalized by the fluid film thickness) predicted for normal continuous approaches in typical AFM and SFA experiments in water (viscosity 0.001 Pa·s). The deformation at the centerpoint ($w$) is calculated at a central fluid film thickness 20nm, well within the range of most conservative surface forces. Solid lines: typical AFM experiments with radius 5 μm, spring constant k= 1 N/m, and approaching velocity v at 0.1, 1, 10, 100, 1000 μm/s. Dashed lines: typical SFA experiments with R = 2 cm, spring constant 100 N/m, motor velocities v at 0.1, 1, 10, 100 nm/s. Red arrow and change in color gradient indicate the increasing of motor velocity.

For rigid surfaces in simple geometries (e.g. sphere-plane, crossed cylinders) the Derjaguin approximation can be used to extract the surface potential $E(h)$ from surface force $F_s = 2\pi R E(h)$. In addition, the hydrodynamic contribution to the total force can be further simplified to the Taylor equation.[25] However, for deformable surfaces, the surface geometry as well as the absolute surface separation, can vary due to fluid pressure. Thus, the knowledge of the surface profile as a function of time is required to decouple the hydrodynamic and surface forces. To do so, one needs to estimate how much deformation is present in the separation range where surface forces are measured. As an example, we present here deformation maps (see Figure 2b) by solving Eqns 1-3 for a continuous approach in the absence of surface forces ($Fs = 0$). In Figure 2b the relative deformation at the center point ($w/h$) is plotted for a range of Young's modulus (half-space) for experimental conditions that are typical of AFM ($R = 5\mu m$, $k = 1$ N/m) or SFA ($R = 2$cm, $k = 100$N/m) during dynamic experiments in water for a central separation $h = 20$ nm. As seen in Figure 2b, significant deformation (of order of the fluid film thickness) can be achieved under many realistic experimental conditions involving compliant materials. Based on the deformation plotted in Figure 2b, for an AFM experiment with drive velocity of $v = 1$ mm/s, we calculate the increase in hydrodynamic force at $h = 20$ nm due to the deformation of a material of $E = 1$MPa to be 17%. Similarly, for a SFA experiment at a drive velocity of $v = 100$ nm/s which is fairly fast consider its larger radius, this increase is in force is 10%, even for $E = 5$MPa material. In addition, deformation brings uncertainties in the determination of the contact position during dynamic AFM measurements. Furthermore, possible additional deformation due to shear near contact, which we did not include in the map of Figure 2b, would amplify further the change in fluid film profile.

3. Experimental validation

A limited number of drainage studies have investigated hydrodynamic forces where one or both of the interacting surfaces are able to deform. In general, these have been performed using atomic force microscopy (AFM) and the surface forces apparatus (SFA).[3,13,14,18-20,32,33] However, due to the difficulty in measuring both the deformation of the surface and the hydrodynamic forces, most studies fit direct force measurements data with theoretical approximations for the spatiotemporal deformed surface geometry. Recently, Kaveh et al. measured the hydrodynamic forces between a colloidal sphere and a soft, elastic polydimethylsiloxane (PDMS) surface using an AFM.[13] To approximate the deformation they used a numerical model to describe the normal hydrodynamic stresses acting on the liquid-solid interface. They showed that upon approach the deformation of the elastic layer leads to a reduction in the hydrodynamic forces due to an increase in the surface separation, facilitating additional drainage out of the gap. Whilst upon retraction, an upward deformation of the elastic layer was found to conversely retard the widening of the gap.

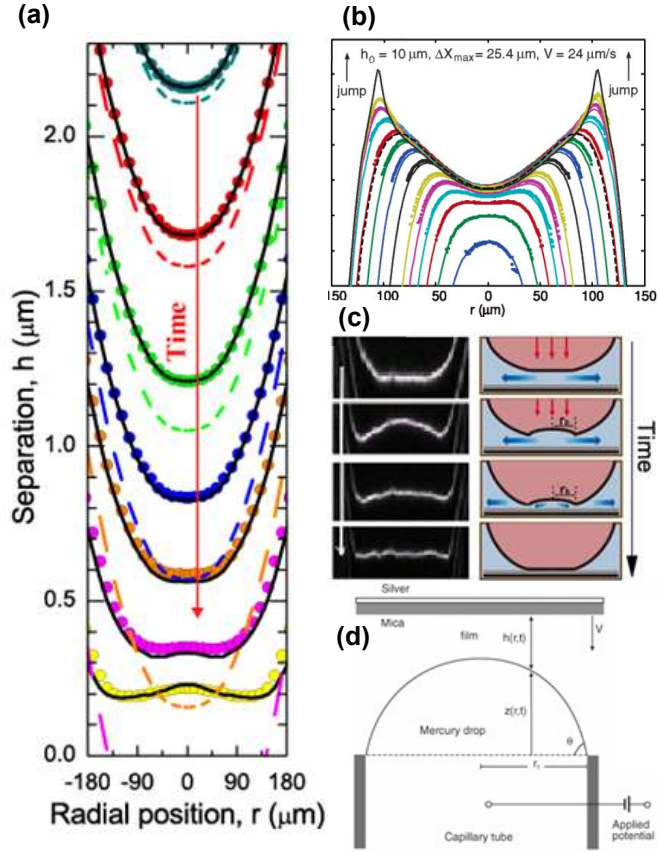

**Figure 3.** Predicted and measured spatiotemporal evolution of the fluid film thickness for (a) an equivalent PDMS elastic sphere approaching a rigid wall in a viscous oil and (b) a mercury droplet with an applied positive potential approaching a negatively charged mica surface in an aqueous electrolyte. In both (a) and (b), the data points represent experimental results and the solid lines are theoretical predictions. In (a) the dashed lines correspond to the theoretical undeformed surface profiles. Schematic diagrams of the experimental configurations used to obtain results in (a) and (b) are shown in (c) and (d), respectively. Figure (a),(c) are reprinted from [14], and (b),(d) are from [17] with permissions.

One possible difficulty in interpreting force measurements is that different contact geometries could potentially result in the same force. Therefore, direct measurements of surface deformation as a function of absolute separation are particularly helpful to understand the role of elastohydrodynamics on dimpling and conservative surface forces. Recently, we demonstrated that the deformation of an elastic surface approaching a rigid surface in a viscous fluid can be directly monitored as a function of absolute separation in a SFA (Figure 3a,c).[14] We showed that the deformation profiles were better described as a viscoelastic solid using the Kelvin-Voigt model than by treating the film simply as an elastic solid. By performing measurements for different PDMS film thicknesses, we also showed that for a thick film EHD leads to the formation of a dimple, preventing the interacting surfaces to make full contact, while for a thinner film contact could be made. This difference in the contact mechanics for the thin film was attributed to the stresses in the film being supported by the underlying substrate. In addition, we found that, despite having the same bulk mechanical properties, the effective modulus of a deformable layered system obtained by static force measurements cannot be applied in the analysis of dynamic interactions.

To avoid contributions from an underlying substrate when trying the measure the properties of elastic thin films, Charlaix and coworkers developed a nanorheological method which monitors the small amplitude oscillation of a sphere close to an elastic surface.[3] Based on the interplay of the liquid viscosity and solid elasticity, they showed that the viscous and elastic contributions to the damping force can be derived from a continuous elastohydrodynamic model (Figure 2a). By applying this technique and analysis, they demonstrated that the surface elastic properties of thin films can be characterized at nanoscale separations without mechanical contact, therefore eliminating the effects of adhesion and preventing surface contamination. In a related study, this technique has also been demonstrated to be a powerful method for studying the surface elasticity and interfacial properties of bubbles or air pockets trapped in hydrophobic structures.[33]

A number of similarities can be drawn between the elastohydrodynamic interactions between soft, solid interfaces and those of gas bubbles or liquid droplets (which have been investigated in more details). For example, Horn and coworkers developed a modified version of the SFA to measure the surface and hydrodynamic forces between a mercury drop and flat solid surface.[34] By analyzing the local curvature of the mercury/aqueous interface and relating it to the pressure drop across the interface via the Young-Laplace equation, they showed that the total film pressure does not vary greatly during the formation of a dimple or the thin film drainage process for different magnitudes and signs of disjoining pressures.[35] Instead, for different electrical potentials between the mercury drop and the aqueous phase, the hydrodynamic pressure was found to adjust to balance the disjoining pressure so that the total film pressure is approximately constant.

More complex flow phenomena and deformation profiles, such as dimples, wimples, pimples and ripples, have also been observed between drops or a drop and a sold surface. These complex geometries have been shown to be similarly well-described by combining the Stokes-Reynolds and Young-Laplace equations.[16,18] Using this analysis it was demonstrated that a repulsive hydrodynamic force leads to the formation of a dimple prior to contact, whereas an attractive double layer force causes a collapse or "jump in" at the dimple barrier rim (Figure 3b,d), where the surface forces are expected to be largest.[17] On the contrary, some repulsive surface forces, such as van der Waals forces, have been shown to further prevent the surfaces from collapsing or coalescing in order to maintain a minimum fluid thickness between them.[21] In a number of related studies, Chan and coworkers have developed a full numerical theory based on this analysis to describe the dynamic interactions of drops and bubbles measured using AFM (Fig. 3b).[16] In doing so, they have defined key parameters to describe the deformation behavior of drops as function of separation.[20] However, although there are many similarities which can be drawn between the dynamic interactions of drops and bubbles with those of soft solids, for drops and bubbles surface tension plays a key role in determining their shape and deformation. In contrast, for soft solids surface tension is expected to be less pronounced, therefore their deformation mechanisms are different. Though it should be noted that elastocapillary effects[36] could also play a role in the case of solid-solid interactions.

In absence of a moving spring which limits the surface movement, many experiments involving the bouncing of a sphere have been conducted to validate the EHD theory. These studies are important in the context of collisions,[28,37,38] as well as in hydrodynamic detachment for applications such as filtration, surface cleaning, and biological adhesion[22] Additional studies investigated the surface interactions and contact behaviors of viscous and viscoelastic thin films. For example, Zeng et al. combined *in situ* microscopy with interferometry in an SFA to measure the separation, film thickness, refractive index and contact geometry of adhesive contacts of viscous liquids and viscoelastic thin films.[39] They showed that the rate of growth of contact area could be well-predicted by a power law for adhesive contact and sintering of two Maxwell viscoelastic spheres. The viscoelastic behavior of confined viscous liquids has also been studied by Villey et al.[24] Their results showed that when confined to nanometer scales, the rheological properties of liquids cannot be decoupled from the global system response.

The confinement of pressurized synovial fluid has been hypothesized to contribute to the lubricating mechanism in mammalian joints by preventing direct contact and therefore avoiding wear-induced failure.[40] Analogous to this behavior, Espinosa-Marzal et al. showed that the confinement-induced pressurization of highly viscous lubricants can lead to the enhancement of the effective elastic modulus of polymer brush surfaces and shield polymer brush layers from the effects of the load.[9] In a related study, Charrault et al. studied via AFM measurements the effects of EHD on polymer brushes, and their effects on the hydrodynamic drainage forces.[41] By shifting the force curves by the length of the unperturbed polymer brush they showed that the hydrodynamic forces of a thick polymer brush layer could be described with a boundary slip model at low velocities (Figure 4). However, at high velocity the hydrodynamic forces were found to deviate from their analysis.

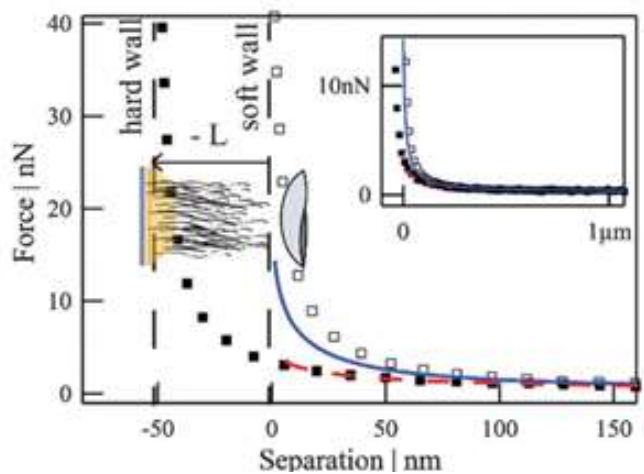

**Figure 4.** Hydrodynamic drainage force (open squares) plotted as a function of separation for a silica microsphere ($R \sim 10$ μm) approaching a thick, end grafted PEG polymer brush. The black squares represent the same data if shifted by the equilibrium brush thickness (50 nm). The solid (blue) line represents the no slip model and the dashed (red) line represents the slip model fitted to the shifted force curve, with a slip length. The inset shows the same force curves and model predictions at larger separations. Figure reprinted from [41] with permission.

A dependence on approach velocity has also been observed in the hydrodynamic interactions of other poroelastic systems. For example, using a novel colloidal probe AFM method with a direct magnetic drive to remove cantilever drag effects, Nalam et al. demonstrated that the load dependence of the storage stiffness of polyacrylamide hydrogels transitioned from Hertzian to a dynamic punch-type model at higher frequencies.[42] This change in the contact mechanical behavior of the hydrogels was attributed to a contribution from solvent confinement or hydrogel poroelasticity, which was found to lead to a stiffening of the mechanical response of the gels.

## 4. Current challenges

**Development of new instrumentation and techniques** In AFM studies, the absolute separation between the interacting surfaces and associated deformation can typically only be obtained indirectly through theoretical modeling. In recent years a number of efforts have been made to integrate the measurement of the interaction force of deformable surfaces with simultaneous spatial-temporal visualization. For example, Tabor et al.[21] combined AFM and laser scanning confocal microscopy to analyze the interactions between fluorinated oil droplets in water. Other groups also combined AFM with reflection interference contrast microscopy (RICM). For instance, Shi et al. measured the interactions between an air bubble and solid surface of different hydrophobicites.[43] However, Erath et al.[44] used this approach to study the quasi-static adhesive-elastic interactions between solids, dynamic AFM-RICM measurements involving soft solid surfaces have yet to be performed.

Different versions of the SFA have also similarly been modified to enable simultaneous optical[39] and fluorescence[45] imaging of the contact area during surface force measurements. In particular, such in-line visualization has been demonstrated to provide valuable additional insight into the adhesion mechanisms of viscoelastic thin films.[39] Other modifications to the SFA have also been implemented by Charlaix and coworkers to allow the measurement of the relative displacement of the surfaces with a capacitance sensor, hence allowing for surface deformation to be uncoupled from the force spring bending. [3,33,46] However, this technique does not allow for direct characterization of the contact shape. On the other hand, our own group has had success with using interferometry to directly monitor the elastohydrodynamic deformation of elastic films, but have had to rely on theoretical models to estimate far-field deformations in the calculations of forces.[14]

In addition to AFM and SFA, many lab-built microtribometers have been developed in the last ten years that combine force measurements with *in situ* imaging and/or in line laser techniques. These have been used to study the adhesive contact geometries[47] or role of surface deformation in the friction of compliant surfaces.[48] However, we are not aware that any such instruments have been employed for dynamic force measurements in fluid environments. In other experimental set-ups, the embedding of fluorescent particles combined with laser profilometry (Figure 5b) has been recently demonstrated to be a useful method to capture the *in situ* deformation of elastic thin films in oil.[49] However, in this form of imaging the spatial resolution of deformation is limited. On the other hand, it has been demonstrated that physical and chemical methods, such as quick quenching with liquid nitrogen and UV cross-linking, can provide a means to capture transient surface patterns and instabilities in viscoelastic thin films for *ex situ* image analysis.[39]

**Viscoelasticity.** The majority of existing studies and theoretical models have focused on accounting for the deformation of elastic contacts. However, many soft materials are inherently viscoelastic. A number of recent studies demonstrated the need to consider the role of viscoelasticity of the interacting materials in the theoretical treatment, and to characterize the role of viscoelasticity on adhesion and contact dynamics. For example, in a recent theoretical study, Pandey et al. presented how the viscoelasticity of a soft wall lifted the sliding of a lubricated sphere (Figure 5a).[50] For microstructured epoxy surfaces, results from Castellanos et al. have demonstrated that viscoelasticity can determine the adhesion properties of structured surfaces more than their bulk elasticity (Young's modulus).[51] Other investigations for smooth surfaces, such as by Zeng et al., have shown existing contact mechanical models to fail for viscous liquids and viscoelastic thin films.[39] Similarly, Nalam et al. observed the Hertzian model to be limited to low frequencies and strains in dynamic modulation measurements on hydrogel networks, further demonstrating the complex and dissipative nature of viscoelastic materials.[42] Analogously, in our own analysis, we also showed that the spatiotemporal deformation of an elastic film for elastic layers in viscous fluid can be best described by treating the elastic film as a viscoelastic material.[14]

**Soft coatings and finite thickness films.** Another complication to EHD, which has been addressed by very limited number studies, is that of layered or stratified materials. Nano-rheological techniques have been shown to be promising to probe the dynamic responses of viscoelastic properties of materials by separating the viscous and elastic contributions to their overall mechanical behaviors. Using this type of approach, Leroy et al. extracted the absolute modulus of an elastic soft layer, in both limits of a thin and thick film, by applying a finite thickness elastic model.[3] Such observations are also supported by our results, which showed that the finite thickness of the elastic film gradually alters the deformation profile from that of a semi-infinite compliant material during the drainage measurements in a viscous oil.[14]

Compared to an elastic half-space, an elastic layer of finite thickness on a rigid substrate would be more general and relevant to applications such as nano-scale devices and thin film coatings. For the case of contact mechanics McGuiggan et al. provided a full numerical solution for indentation (contact) of multilayer system. To verify their model they compared their numerical results to the contact radii and forces measured for the same system using an SFA.[52] In the case of EHD, the fluid pressure distribution is not the same as in contact mechanics. Therefore, corrections based on a contact mechanics treatment cannot be applied for EHD of stratified materials.[14] Modeling general elasticity solutions in 3D systems is mathematically complex and analytical solutions for only a few simple geometries are attainable. Instead, a more basic approach one can take is to consider only a 2D geometry, for example in axisymmetrical situations. This theoretical analysis has been previously employed by Balmforth et al. who studied the sedimentation of objects under gravity toward a layered elastic system.[53] By solving this EHD problem they compared the different contact profiles for a few limiting cases: as a foundation (thin compressible layer on rigid base), half-space, beam and membrane, for varying thicknesses of the elastic layer. In these 2D limits of very thin films they predicted how the asymptotic scaling of a central and barrier rim fluid gap thickness changes.

**Path-dependent deformation.** The dissipative nature of fluid drainage processes leads to surface deformation and fluid flow during EHD that depends on the history of the surface motion. Such path-dependent deformation profiles were demonstrated by Clasohlm et al., who investigated the sudden approach of a solid wall towards a liquid drop from close proximity.[18] Using a modified SFA to monitor the contact geometry and absolute surface separation, they showed that unusual shapes of the mercury drops could occur depending on the approach conditions. In particular, they observed the formation of a wimple en route to an eventual dimple, which was not observed for approaches from large initial separations. It can be anticipated that such path dependent flow and contact geometries would also occur for a soft solid. However, to the best of our knowledge, no previous study has yet to discuss this possibility in detail.

**Slip.** In general, the no-slip boundary condition has been used to describe the fluid flow at wetting interfaces. However, the validity of available corrections to the no-slip boundary condition have yet to be investigated systematically for deformable surfaces. Furthermore, the incorporation of slip corrections to a deformable surface is not straightforward.[54] For a polymer brush layer, Charrault et al. demonstrated that the hydrodynamic drainage of a fluid squeezed between a sphere and a thick brush could be well described by the invoking a slip length, if the surface separation was corrected for a non-deformed zero position (Figure 4).[41] However, for a large approach velocity they observed a notable deviation from the separation-corrected slip model which they attributed to prior contact elastic deformation of polymer brushes due to the larger magnitude of the repulsive hydrodynamic force at a high velocity.

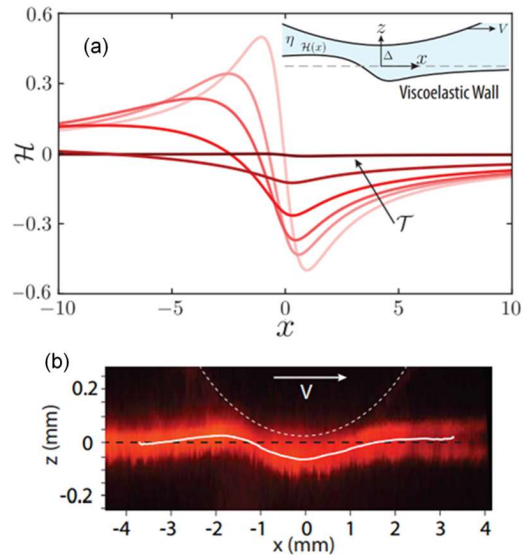

**Figure 5.** (a) Theoretical prediction for the deformation of a viscoelastic wall during the sliding of a standard linear solid half space (shown in inset) for various characteristic time scale $\mathcal{T}$, where $\mathcal{T} = 0$ signifies a purely elastic wall. Figure reprinted from [50](b) Laser profilometry image showing the deformation of the elastic thin film with fluorescent particles embedded during sliding of a negatively-buoyant rigid cylinder immersed in a viscous oil bath on a tilted elastic wall. Figure reprinted from [49] with permission.

**Heterogeneous (porous) materials.** Difficulties in modelling EHD are also met when dealing with heterogeneous surfaces, such as porous or structured surfaces. It is possible that compliant materials can be poroelastic. In that case the EHD models, as described in the previous section, must be greatly modified to account for material poroelasticity as the boundary condition on a porous interface is more complex and needs to be carefully considered to describe the coupling between the deformation of solid skeleton and flow of permeating fluid which do not exist in purely elastic solids.[2] The hydrodynamics of poroelastic ordered and disordered microstructures have been studied by Gopinath et al.[55]. Instead of using traditional Brinkman approximations for porous media, they demonstrated that to fully consider the flow through pores a modified boundary condition must be invoked, adding further complexity.

**Conclusions**

We have reviewed a number of important experimental and theoretical studies which have investigated the effects of surface deformation on hydrodynamic forces. By and large, previous experimental studies have been performed using AFM or SFA. However, typically these techniques need to rely on numerical models to access information about the deformation profile or geometry. To facilitate the direct visualization of surface deformation with force measurements a number of new techniques are emerging, for example RICM[43] and laser profilometry. [49]

For rigid surfaces, hydrodynamic interactions in a low Reynolds number Newtonian fluid have been well described by Reynolds equation in the lubrication limit. Conversely, in the case of soft surfaces the deformation of the interface will alter the hydrodynamic interactions. For elastic surfaces, the linear elasticity model provides simple forms of deformation for a half-space and needs to be coupled with the lubrication equation. [28] However, due to finite thickness effects, the deformation for layered materials is more complex and needs to be studied further.

In addition, we discussed more complicated scenarios where the material poroelasticity, viscoelasticity or physical or chemical heterogeneities, such as slip, need to be considered. These effects are particularly important in understanding the role of hydrodynamic forces in biological systems, for instance in wet bio-adhesion. A better understanding in EHD can be anticipated to aid the future development of biomedical implants, such as artificial organs or tubing[7], as well as microfluidic sensors and devices. Furthermore, elastohydrodynamic interactions may provide an additional avenue to fabricate complex 3D structures, such as self-folded structures from 2D planar surfaces and allow more precise control of capillary interactions in micro-electromechanical systems (MEMS). Together, these broader aspects of applications provide a strong motivation for the future study of hydrodynamic interactions in the presence of deformable surfaces.

## Acknowledgements

This work was partially supported by the Donors of the American Chemical Society Petroleum Research Fund under Grant 51803-ND5, the Hopkins Extreme Materials Institute (HEMI), and NSF-CMMI 1538003.
## References

1. Lugt P, Morales-Espejel GE: **A review of elasto-hydrodynamic lubrication theory.** *Tribology Transactions* (2011) **54**(3):470-496.

2. Skotheim JM, Mahadevan L: **Soft lubrication: The elastohydrodynamics of nonconforming and conforming contacts.** *Physics of Fluids* (2005) **17**(9):092101.

3. Leroy S, Charlaix E: **Hydrodynamic interactions for the measurement of thin film elastic properties.** *J Fluid Mech* (2011) **674**(389-407).

4. Snoeijer J, Eggers J, Venner C: **Similarity theory of lubricated hertzian contacts.** *Physics of Fluids (1994-present)* (2013) **25**(10):101705.

5. Urzay J: **Asymptotic theory of the elastohydrodynamic adhesion and gliding motion of a solid particle over soft and sticky substrates at low reynolds numbers.** *J Fluid Mech* (2010) **653**(391-429.

6. Vlassopoulos D, Cloitre M: **Tunable rheology of dense soft deformable colloids.** *Current Opinion in Colloid & Interface Science* (2014) **19**(6):561-574.

7. Jin Z, Dowson D: **Elastohydrodynamic lubrication in biological systems.** *Proceedings of the Institution of Mechanical Engineers, Part J: Journal of Engineering Tribology* (2005) **219**(5):367-380.

8. Dirks J-H, Federle W: **Fluid-based adhesion in insects–principles and challenges.** *Soft Matter* (2011) **7**(23):11047-11053.

9. Espinosa-Marzal RM, Bielecki RM, Spencer ND: **Understanding the role of viscous solvent confinement in the tribological behavior of polymer brushes: A bioinspired approach.** *Soft Matter* (2013) **9**(44):10572-10585.

10. Greene GW, Banquy X, Lee DW, Lowrey DD, Yu J, Israelachvili JN: **Adaptive mechanically controlled lubrication mechanism found in articular joints.** *Proceedings of the National Academy of Sciences* (2011) **108**(13):5255-5259.
17